\documentclass[preprint]{revtex4}%
\usepackage{amsmath}
\usepackage{amsfonts}
\usepackage{amssymb}
\usepackage{graphicx}
\usepackage{epsf,color,colordvi,pifont}
\usepackage{amsfonts}%
\providecommand{\U}[1]{\protect\rule{.1in}{.1in}}
\begin{document}
\title{Noncommutative magnetic moment of charged particles}
\author{T.C. Adorno$^{1}$, D.M. Gitman$^{1}$, A.E. Shabad$^{2}$, D.V.
Vassilevich$^{3,4}$}
\affiliation{$^{1}$Instituto de F\'{\i}sica, Universidade de S\~{a}o Paulo, Brazil}
\affiliation{$^{2}$P.N.~Lebedev Physics Institute, Moscow, Russia}
\affiliation{$^{3}$CMCC - Universidade Federal do ABC, Santo Andr\'e, S.P., Brazil}
\affiliation{$^{4}$Department of Physics, St.~Petersburg State University, Russia}

\begin{abstract}
It has been argued that in noncommutative field theories\textbf{,} sizes of
physical objects cannot be taken smaller than an \textquotedblleft elementary
length\textquotedblright\ related to noncommutativity parameters. By
gauge-covariantly extending field equations of noncommutative $U(1)_{\star}%
$-theory to cover the presence of external sources, we find electric and
magnetic fields produced by an extended static charge. We find that such a
charge, apart from being an ordinary electric monopole, is also a magnetic
dipole. By writing off the existing experimental clearance in the value of the
lepton magnetic moments for the present effect, we get the bound on
noncommutativity at the level of $10^{4}\mathrm{TeV}$.

\end{abstract}
\maketitle

The noncommutative (NC) field theory suggests a very profound revision of the
idea of space and time by referring to 4-coordinates $X^{\mu}$ as operatorial,
noncommuting quantities, $[X^{\mu},X^{\nu}]=i\theta^{\mu\nu}$. Usually the
antisymmetric NC tensor $\theta^{\mu\nu}$ is taken as constant and small in
its magnitude; this will be our choice, too. Due to uncertainty relation
intrinsic to the noncommutativity \cite{DFR}, various components of the
coordinate 4-vector cannot be simultaneously given definite values. This
implies that sizes of physical objects in this theory cannot be taken smaller
than an \textquotedblleft elementary length\textquotedblright. Throughout this
paper we consider the space-space noncommutativity. This means that a
reference frame \cite{footnote} is admitted to exist, wherein $\theta^{0\nu
}=0$, so that the remaining NC parameters can be combined in the 3-vector
$\theta^{i}\equiv(1/2)\varepsilon^{ijk}\theta^{jk}$ in that frame.

The characteristic length is defined through the absolute value of this vector
as $l_{\mathrm{NC}}^{2}=|\theta|\equiv(\hbar c/\Lambda_{\mathrm{NC}})^{2}$,
where $\Lambda_{\mathrm{NC}}$ is the corresponding energy scale. Hence, all
the sources should not be point-like, but rather have a characteristic size of
the order of $l_{\mathrm{NC}}$, see also Ref.\ \cite{SmSp}.

The aim of this work is to study the field produced by a finite-size static
charge in NC electrodynamics. To treat such charges we need to avoid the
difficulty caused by the gauge-invariance violation by a classical external
source, analogous to the trouble encountered in non-Abelian field theories
\cite{SikWei78}. This will be achieved by extending the Seiberg-Witten (SW)
map \cite{SeiWit99} to the case when external currents are present in the
lowest nontrivial order with respect to the NC parameters. With this extension
in hands, we find corrections to the electromagnetic potentials produced by a
finite-size static electric charge. Solutions, regular everywhere, neither can
nor should have a point-charge limit. By selecting such solutions we
essentially part from the standard commutative case. We find that a static
electric charge $eZ$ distributed in a spherically-symmetric way over a sphere
of a finite radius $a$, apart from being an ordinary electric monopole, is
also a magnetic dipole. Its magnetic moment is directed along the NC vector
$\boldsymbol{\theta}$ and its value is quadratic in the charge $eZ$ and
depends on the size $a$ of the latter. In this way we define the NC
contribution to the magnetic moment of an elementary particle viewed upon as a
classical particle with its electric charge distributed according to the
electromagnetic form-factor. This NC contribution appears to be proportional
to $l_{\mathrm{NC}}^{2}/a$, with $a$ being the charge radius. Then, a
comparison with experimental results allows us to establish restriction on
$l_{\mathrm{NC}}$ (or on $\Lambda_{\mathrm{NC}}$). The strongest bounds are
coming from the measurements of the anomalous magnetic moments of leptons
under the assumption that the charge radius is given by the noncommutativity
length, $a\sim l_{\mathrm{NC}}$.

Previously, an NC magnetic solution for the field of a static electric charge
was found by Stern \cite{Stern}, who, in contrast to our work, assumed that
the charge is truly point-like (of zero radius). The results then differ
drastically from ours, and we shall present a comparison between the two approaches.

As an NC space we take the Moyal plane equipped with the Moyal star product
$\check{f}\left(  x\right)  \star\check{g}\left(  x\right)  =\check{f}\left(
x\right)  \exp\left[  \left(  i/2\right)  \overleftarrow{\partial}_{\mu}%
\theta^{\mu\nu}\overrightarrow{\partial}_{\nu}\right]  \check{g}\left(
x\right)  $. We refer to the non-Abelian action of an NC ${U}(1)_{\star}$
gauge theory $\check{S}=\check{S}_{\mathrm{A}}+\check{S}_{\mathrm{jA}}$,%
\begin{equation}
\check{S}_{\mathrm{A}}=-\frac{1}{16\pi c}\int dx\check{F}_{\mu\nu}\star
\check{F}^{\mu\nu}\,,\ \ \check{S}_{\mathrm{jA}}=-\frac{1}{c^{2}}\int
dx\check{j}^{\mu}\star\check{A}_{\mu}\,\label{theory}%
\end{equation}
that consists of the standard gauge-invariant part $\check{S}_{\mathrm{A}}$,
where $\check{F}_{\mu\nu}=\partial_{\mu}\check{A}_{\nu}-\partial_{\nu}%
\check{A}_{\mu}+ig\left[  \check{A}_{\mu}\overset{\star}{,}\check{A}_{\nu
}\right]  ,$ and the part $\check{S}_{\mathrm{jA}}$ responsible for the
interaction of the electromagnetic field potential $\check{A}_{\mu}$ with an
external current $\check{j}^{\mu}$. Here and in what follows the designation
$\left[  \overset{\star}{,}\right]  $ means the Moyal commutator, while the
gauge coupling constant\textbf{ }$g$\textbf{ }is, as usual, identified with
the elementary electric\textbf{ }charge $g=e/(\hslash c)$ (see e.g. Guralnik
\textit{et. al. }\cite{SW-app})\ in order that the interaction strength
between the electromagnetic and a complex, say, spinor field $\check{\psi}$
might be fixed in a gauge-invariant way as $\int dx\,\check{\bar{\psi}}%
\star\gamma^{\mu}\left(  \partial_{\mu}-ie/(\hslash c)\check{A}_{\mu}%
\star\right)  \check{\psi}$. We shall still be destinguishing the constants
$e$ and $g$ until their explicit mutual identification is needed. The
compatibility condition $\check{D}_{\mu}\delta\check{S}/\delta\check{A}_{\mu
}=0$ of the equations of motion $\delta\check{S}/\delta\check{A}_{\mu}=0$
requires that the current and the field be related by the equation of
covariant current-conservation $\check{D}_{\mu}\check{j}^{\mu}=\partial_{\mu
}\check{j}^{\mu}+ig[\check{A}_{\mu}\overset{\star}{,}\check{j}^{\mu}]=0$. This
cannot provide the vanishing of the variation $\delta\check{S}_{\mathrm{jA}%
}=-(1/gc^{2})\int dx\left\{  \left(  \partial_{\mu}\check{j}^{\mu}\right)
\star\check{\lambda}\right\}  $ under a gauge transformation with the
parameter $\check{\lambda}$, because it would require the conservation law
$\partial_{\mu}\check{j}^{\mu}=0$, incompatible with the equations of motion,
see \cite{AGSV}. Hence, the total action $\check{S}$ is not gauge-invariant.
To handle this difficulty we shall in what follows be basing on the field
equation $\delta\check{S}/\delta\check{A}_{\mu}=0$, which is gauge-covariant.

The $U(1)_{\star}$ gauge theory is consistent as it satisfies the criteria of
Ref. \cite{gauge}. Therefore, it is not necessary to consider SW map or to
make the gauge transformations twisted \cite{twist}. However, for studying
phenomenological aspects of an NC theory it is advisable \cite{SW-app} to
perform the SW map, since it allows one to work with commuting electromagnetic
fields $A^{\mu}$ that have standard $U\left(  1\right)  $ gauge transformation
properties. It is known that in the lowest nontrivial approximation in the NC
parameter $\theta$, to which\textbf{ }approximation\textbf{ }we\textbf{ }shall
henceforth restrict ourselves, the field $\check{A}_{\mu}$ is SW-mapped as
\begin{equation}
\check{A}_{\mu}=A_{\mu}+\frac{g}{2}\theta^{\alpha\beta}A_{\alpha}\left[
\partial_{\beta}A_{\mu}+f_{\beta\mu}\right]  \,,\label{sw4}%
\end{equation}
where $f_{\mu\nu}=\partial_{\mu}A_{\nu}-\partial_{\nu}A_{\mu}$. In our case
eq. (\ref{sw4}) should be supplemented by the SW map for currents
\cite{Ban,AGSV}
\begin{equation}
\check{j}^{\mu}=j^{\mu}+g\theta^{\alpha\beta}A_{\alpha}\partial_{\beta}j^{\mu
}\,,\label{sw4.2}%
\end{equation}
that is deduced from the requirement that the external current should
gauge-transform covariantly $\delta\check{j}^{\mu}=i\left[  \check{\lambda
}\overset{\star}{,}\check{j}^{\mu}\right]  ,$ the same as the current of
charged particles, e.g. $\check{\bar{\psi}}\gamma_{\mu}\check{\psi},$
does\textbf{. }The SW map is not unique, but one can show \cite{AGSV}, that
the corresponding ambiguity does not affect corrections to the potential of a
static charge to the first order in $\theta$. After the SW map (\ref{sw4}),
(\ref{sw4.2}) is applied to the equations of motion $\delta\check{S}%
/\delta\check{A}_{\mu}=0$ and $\check{D}_{\mu}\delta\check{S}/\delta\check
{A}_{\mu}=0$ one gets the nonlinear field equations with external current,
valid to the first order in $\theta^{\mu\nu},$%
\begin{align}
\partial_{\nu}f^{\nu\mu}-g\theta^{\alpha\beta}\left[  \partial_{\nu}%
(f_{\alpha}^{\ \nu}f_{\beta}^{\ \mu})-f_{\nu\alpha}\partial_{\beta}f^{\nu\mu
}-A_{\alpha}\partial_{\beta}\left(  \partial_{\nu}f^{\nu\mu}-\frac{4\pi}%
{c}j^{\mu}\right)  \right]   &  =\frac{4\pi}{c}j^{\mu}\,,\nonumber\\
\partial_{\mu}j^{\mu}+g\theta^{\alpha\beta}\left(  f_{\mu\alpha}%
\partial_{\beta}j^{\mu}+A_{\alpha}\partial_{\beta}\partial_{\mu}j^{\mu
}\right)   &  =0\,.\label{SWeq}%
\end{align}
The explicit presence of potentials in (\ref{SWeq}) may look disturbing, but
this difficulty is easily solved. To restore covariance we consider a
perturbative solution of equations (\ref{SWeq})\textbf{ }by expanding it in
the noncommutative parameter. Explicitly, starting with the zeroth
approximation $A^{(0)}$, $j^{(0)}$ that satisfies the standard Maxwell
$\partial_{\nu}f^{\left(  0\right)  \nu\mu}=\left(  4\pi/c\right)  j^{(0)\mu}$
and current-conservation $\partial_{\mu}j^{(0)\mu}=0$ equations, we obtain for
the first-order corrections $A^{(1)}$, $j^{(1)}$
\begin{align}
&  \partial_{\nu}f^{(1)\nu\mu}-g\theta^{\alpha\beta}\left(  \partial_{\nu
}(f_{\alpha}^{(0)\nu}f_{\beta}^{(0)\mu})-f_{\nu\alpha}^{(0)}\partial_{\beta
}f^{(0)\nu\mu}\right)  =\frac{4\pi}{c}j^{(1)\mu}\,,\nonumber\\
&  \partial_{\mu}j^{(1)\mu}+g\theta^{\alpha\beta}f_{\mu\alpha}^{(0)}%
\partial_{\beta}j^{(0)\mu}=0\,.\label{SWeq1}%
\end{align}

In what follows we shall study solutions to (\ref{SWeq1}) produced by a static
spherically symmetric charge distribution. It is defined in two regions,
$\mathrm{I:\ }r<a$ and $\mathrm{II:\ }r>a$,$\ r=|\mathbf{x}|,$%
\begin{equation}
j^{(0)\mu}=\left(  c\rho,0\right)  \,,\ \ \rho_{\mathrm{I}}\left(
\mathbf{x}\right)  =\frac{3}{4\pi}\frac{Ze}{a^{3}}\,,\ \ \rho_{\mathrm{II}%
}\left(  \mathbf{x}\right)  =0\,.\label{10.1}%
\end{equation}
The uniform charge distribution inside the sphere, whose radius is $a$, is
taken for simplicity. Extensions to arbitrary spherical symmetric
distributions, continuous ones included, may be also considered, when
necessary. The charge density (\ref{10.1}) tends to the delta-function in the
point-charge limit: $\rho(\mathbf{x})=Ze\delta^{3}(\mathbf{x})$, as
$a\rightarrow0$. We shall argue, however, that the corresponding point-source
solution (the Green function) does not exist even as a standard generalized
function. Once no spherical \textit{physical object} should be taken with its
radius smaller than the elementary length, we will restrict our consideration
to the values $a>l_{\mathrm{NC}}$.

We use the Coulomb gauge $\partial_{i}A^{i}=0$ for the stationary solutions,
to which we confine our consideration. Then the standard Maxwell equations
provide the following spherically symmetric, $A^{\left(  0\right)  \mu}\left(
\mathbf{x}\right)  =A^{\left(  0\right)  \mu}\left(  r\right)  $,
electromagnetic potential $A^{\left(  0\right)  \mu}=\left(  A^{\left(
0\right)  0},0\right)  $,%
\begin{equation}
A_{\mathrm{I}}^{\left(  0\right)  0}\left(  r\right)  =-\frac{Ze}{2a^{3}}%
r^{2}+\frac{3}{2}\frac{Ze}{a}\,,\ \ A_{\mathrm{II}}^{\left(  0\right)
0}\left(  r\right)  =\frac{Ze}{r}\,,\label{10.2}%
\end{equation}
which satisfies the smoothness conditions $\left.  A_{\mathrm{I}}%
^{0}(r)\right\vert _{r=a}=\left.  A_{\mathrm{II}}^{0}(r)\right\vert _{r=a}$,
$\left.  \partial_{r}A_{\mathrm{I}}^{0}(r)\right\vert _{r=a}=\partial
_{r}\left.  A_{\mathrm{II}}^{0}(r)\right\vert _{r=a}$ at the boundary of the
sphere, is regular in the origin $A^{0}\left(  0\right)  \neq\infty,$ and
falls off at infinity $\left.  A^{0}\left(  r\right)  \right\vert
_{r\rightarrow+\infty}=0$.

The analysis presented above is valid for arbitrary constant $\theta^{\mu\nu}%
$. Henceforward we restrict ourselves to the space-space noncommutativity
$\left(  \theta^{0\mu}=0\right)  $. Due to the spherical symmetry and\textbf{
}to the stationarity, the second equation in (\ref{SWeq1}) is satisfied by
$j^{\left(  1\right)  \mu}=0$, no correction to the current is required. This
implies that the current remains dynamically intact, $j^{\mu}=j^{\left(
0\right)  \mu}$, so we may refer to it as a fixed external current, as this is
customary in an $U(1)$-theory. The NC Maxwell equation (\ref{SWeq1}) for the
zeroth component $\left(  \mu=0\right)  $ now reduces to $\mathbf{\nabla}%
^{2}A^{\left(  1\right)  0}=0$\thinspace, so that there is no first-order
corrections $A^{\left(  1\right)  0}\left(  \mathbf{x}\right)  $\ to the
potential, that would satisfy the same boundary conditions. (Such corrections
appear, if a background magnetic field is added to the zero-order solution
(\ref{10.2}), see \cite{AGSV}). However, for the spatial components $\left(
\mu=k=1,2,3\right)  $ we obtain the inhomogeneous Laplace equations%
\begin{align}
&  \mathbf{\nabla}^{2}A_{\mathrm{I}}^{\left(  1\right)  k}\left(
\mathbf{x}\right)  =-g\left(  \frac{Ze}{a^{3}}\right)  ^{2}\theta^{ik}%
x^{i}\,,\nonumber\\
&  \mathbf{\nabla}^{2}A_{\mathrm{II}}^{\left(  1\right)  k}\left(
\mathbf{x}\right)  =-g\left(  \frac{Ze}{r^{3}}\right)  ^{2}\theta^{ik}%
x^{i}\,.\label{es2}%
\end{align}
Their only smooth solution, regular in $r=0$ and decreasing for $r\rightarrow
\infty$ is
\begin{align}
&  A_{\mathrm{I}}^{\left(  1\right)  k}\left(  \mathbf{x}\right)  =-\frac
{g}{4}\left(  \frac{Ze}{a^{2}}\right)  ^{2}\left(  \frac{2}{5}\frac{r^{2}%
}{a^{2}}-1\right)  \theta^{ik}x^{i}\,,\nonumber\\
&  A_{\mathrm{II}}^{\left(  1\right)  k}\left(  \mathbf{x}\right)  =\frac
{g}{4}\left(  \frac{Ze}{r^{2}}\right)  ^{2}\left(  \frac{8}{5}\frac{r}%
{a}-1\right)  \theta^{ik}x^{i}\,.\label{es5}%
\end{align}
This solution neither has nor should have the point-source limit at
$a\rightarrow0$. The leading long-distance part of the vector-potential
$\mathbf{A}_{\mathrm{II}}^{\left(  1\right)  }$ behaves like that of a
magnetic dipole, the static charge (\ref{10.1}) being thus a carrier of an
equivalent magnetic moment $\boldsymbol{\mathcal{M}},$%
\begin{equation}
\boldsymbol{A}=\frac{\left[  \boldsymbol{\mathcal{M}}\times\mathbf{x}\right]
}{r^{3}},\qquad\boldsymbol{\mathcal{M}}=\boldsymbol{\theta}(Ze)^{2}\frac
{2g}{5a}\,.\label{magnmoment}%
\end{equation}

Let us study the consequences of this relation for particle physics. Lower
bounds on the NC scale based on high-energy experiments have been drastically
improved during recent years. The analysis of primordial nucleosynthesis data
\cite{Horvat:2009cm} gives $\Lambda_{\mathrm{NC}}\gtrsim3\,{\text{TeV}}$ as a
conservative estimate, while with other choices of parameters the bound
increases to $10^{3}\,$TeV. From ultra-high energy cosmic ray experiments one
deduces \cite{Horvat:2010sr} that $\Lambda_{\mathrm{NC}}\gtrsim
200\,{\text{TeV}}$. The data for photon-neutrino interaction put the bound for
time-space NC into approximately the same range \cite{Horvat:2011iv}. A much
stronger bound, $\Lambda_{\mathrm{NC}}\gtrsim5\cdot10^{11}{\text{TeV}}$, was
obtained \cite{MPR} by analyzing an atomic magnetometer experiment
\cite{Berglund}. Characteristic energy scale of this experiment is below
$1\,{\text{eV}}$, while the typical scale of modern particle physics
experiments reaches TeV. Between these scales the value of effective NC
parameter may change considerably. One of mechanisms for such a change may be
due to the QFT effects which may manifest themselves through the
renormalization group variation of couplings with characteristic energy.
Therefore, it is important to study independently the high-energy
restrictions, which we do below by using (\ref{magnmoment}).

For a particle of unit charge, $Z=1$, and mass $m$ the NC correction to
magnetic moment reads
\begin{equation}
\delta_{\mathrm{NC}}|\boldsymbol{\mathcal{M}}|=\alpha|\theta|\mu\frac{4m}%
{5a}\,,\label{delM}%
\end{equation}
where $\alpha$ is the fine structure constant and $\mu=e/(2m)$ is the
corresponding magneton. From now on, we put $\hbar=c=1$ and, consequently,
$g=e$. For the proton, by taking the charge radius of $0.9\,\mathrm{fm}$ for
$a$ we conclude that the correction to the magnetic moment is below the
experimental error of $2.3\cdot10^{-8}\mu_{N}$ \cite{PDG} already for
$\Lambda_{\mathrm{NC}}\simeq0.24\,\mathrm{TeV}$. An estimate of the NC proton
magnetic moment contribution into the hyperfine splitting of the energy states
in a hydrogen atom, based on a NC theory \cite{gauge, ABCGT} of electron
spectrum, does not strengthen the bound on $l$ found in \cite{Stern} with the
use of the non-dipole magnetic solution given as eq.(\ref{15}) below\textbf{.}
In the case of leptons, we require that NC corrections to the magnetic moment
anomaly,
\begin{equation}
\delta_{\mathrm{NC}}((g_{l}-2)/2)=4\alpha|\theta|m_{l}/(5a_{l})\label{delg}%
\end{equation}
lie within experimental errors, which are $3\cdot10^{-13}$ for electrons, and
$6\cdot10^{-10}$ for muons \cite{PDG}. With the estimate $a_{e},a_{\mu
}<10^{-3}fm$ that corresponds to the LEP energy scale of 200 Gev we obtain
$\Lambda_{\mathrm{NC}}\gtrsim45\,\mathrm{TeV}$ in the case of electrons and
$\Lambda_{\mathrm{NC}}\gtrsim14\,\mathrm{TeV}$ in the case of muons. These two
bounds are of the order of currently accepted restrictions on the NC scale,
but do not improve them.

The situation changes if we accept that the charge radius of leptons is
defined solely by the NC effects. That is, $a_{e}\simeq a_{\mu}\simeq
\sqrt{|\theta|} =\Lambda_{NC}^{-1}$. Then, from the restrictions on the muon
anomalous magnetic moment we derive $\Lambda_{\mathrm{NC}}\gtrsim10^{3}\,
\mathrm{TeV}$, while for the electron we have $\Lambda_{\mathrm{NC}}%
\gtrsim10^{4}\, \mathrm{TeV}$, or $l_{\mathrm{NC}}\lesssim2\cdot
10^{-8}\mathrm{fm}$. This is the strongest bound on the NC scale among the
ones, which follow from the high-energy data.

In the argumentation above we used the experimental errors only while
completely ignoring possible theoretical uncertainties. This can be done for
the following simple reason. Any theoretical calculation based on the usual
commutative quantum field theory predicts the magnetic dipole moment directed
along the spin vector ${\boldsymbol{S}}$, which characterizes the state of a
particle. The NC correction (\ref{magnmoment}) is parallel to the NC vector
$\boldsymbol{\theta}$, which is a characteristic of the background space-time.
We expect that relative orientation of $\boldsymbol{S}$ and
$\boldsymbol{\theta}$ taken for various particles in various experiments is
random. The effect of noncommutativity is in widening the range of
experimental data rather than in shifting the central value. Therefore, the
experimental error does give a bound on the NC effects even without taking
into account theoretical uncertainties.

Yet another, also smooth, solution of equation (\ref{es2}) for the
vector-potential is worth discussing
\begin{align}
A_{\mathrm{I}}^{\left(  1\right)  k}\left(  \mathbf{x}\right)   &  =-\frac
{g}{4}\left(  \frac{Ze}{a^{2}}\right)  ^{2}\left(  \frac{2}{5}\frac{r^{2}%
}{a^{2}}+\frac{8}{5}\frac{a^{3}}{r^{3}}-1\right)  \theta^{ik}x^{i}%
\,,\nonumber\\
A_{\mathrm{II}}^{\left(  1\right)  k}\left(  \mathbf{x}\right)   &  =-\frac
{g}{4}\left(  \frac{Ze}{r^{2}}\right)  ^{2}\theta^{ik}x^{i}\,,\label{15}%
\end{align}
that is not regular in the origin, but decreases at large distance from the
source faster than (\ref{es5}), in other words it is more localized. Unlike
eq. (\ref{es5}) solution (\ref{15}) is not the field of a magnetic dipole,
since it decreases away from the source faster than that. The second line in
(\ref{15}) does not depend on the size $a$ of the charge and coincides with
the magnetic solution found in \cite{Stern} for the field produced by a
point-like static charge outside of it, i.e. for $r\neq0$. It is highly
singular, $\sim1/r^{3}$, in the origin $r=0$. Correspondingly, it does not
make a solution in a reasonable class of generalized functions, when continued
to the point $r=0$. (In this respect it deeply differs from the standard
solution $A_{\mathrm{II}}^{\left(  0\right)  0}$ in (\ref{10.2}), which, in
the limit $a\rightarrow0$, is less singular, $\sim1/r$, and makes a
generalized-function solution to the Laplace equation with $\delta
^{3}(\mathbf{x})$ as its inhomogeneity (see \textit{e.g}. \cite{Vlad}). That
solution is defined in the whole $\mathbb{R}^{3}$, the point $r=0$ included.)
For this reason our choice is in favor of the nonsingular solution (\ref{es5}).

One can consider the solution which satisfies weaker conditions at infinity,
so that an external homogeneous magnetic field is allowed. In such a case, one
can find \cite{AGSV} many interesting similarities and differences with the
QED effects \cite{ShaUs07}.

To summarize, our main result is the (remarkably simple) formula
(\ref{magnmoment}) for NC magnetic moment of a spherical charge.\textbf{ }Eq.
(\ref{magnmoment}) is subject to quantum corrections and classical corrections
of higher powers in the noncommutativity parameter $\theta$, which are both
small. So, we are confident that this result will remain valid in a more
complete approach, as well as the bounds it imposes\textbf{ }on the NC scale.

We are grateful to Josip Trampetic for explaining to us experimental bounds on
the NC scale. This work was supported by FAPESP, CNPq and by RFBR under the
Project 11-02-00685-a.


\begin{thebibliography}{99}
\bibitem {footnote}This frame can be, for instance, associated with the rest
frame of the Earth surface.
\bibitem {SmSp}A.~Smailagic, E.~Spallucci,
J.\ Phys.\ A \textbf{A36}, L517-L521 (2003). In this paper it is suggested
that the smearing of the sources is a substitute for the noncommutative
multiplication, so that the product between fields in the classical action
remains the usual commutative one. We \emph{do not} adopt this point of view.
\end{thebibliography}
\end{document}